\documentclass{aa}
\usepackage{graphics}

\begin{document}
\thesaurus{08(09.02.1;09.07.1;09.08.1;11.09.1 NGC\,55;11.09.4)}
\title{Long slit spectroscopy of diffuse ionized gas in NGC\,55\thanks{Based on
observations obtained at ESO/La Silla}}
\author{B. Otte\inst{1,2} \and R.-J. Dettmar\inst{1}}
\institute{Astronomisches Institut, Ruhr-Universit\"at Bochum, D-44780 Bochum,
Germany \and Department of Astronomy, University of Wisconsin, Madison, WI
53706, USA}
\mail{otte@astro.wisc.edu}
\date{Received 26 October 1998; accepted 4 January 1999}
\titlerunning{Spectroscopy of DIG in NGC\,55}
\maketitle

\begin{abstract}
Spectroscopic measurements of emission line ratios and velocities are presented
for ionized gas across the central plane of NGC\,55, a late-type galaxy in the
Sculptor group. The low metallicity in NGC\,55 leads to relatively low line
ratios of [\ion{S}{ii}]/H$\alpha$ and [\ion{N}{ii}]/H$\alpha$ for \ion{H}{ii}
regions as well as the diffuse ionized gas. These are the first spectroscopical
measurements of line ratios in ionized gas exterior to the stellar disc of
NGC\,55. Analysis of the line ratios and the relative velocities of different
features suggests that photoionization is a plausible explanation for the
ionization of this diffuse gas. The observed shell structures and the
corresponding velocities support the idea of diffuse gas being pushed into the
halo by supernova explosions and stellar winds.
\end{abstract}

\section{Introduction}

When examining ionized gas, it is common practice to distinguish between
classical \ion{H}{ii} regions (Str\"omgren spheres around OB stars) and diffuse
ionized gas (DIG), the gas outside the boundaries of the Str\"omgren spheres.
While \ion{H}{ii} regions are created by photoionization, the ionization
processes for the diffuse gas are less well known. Many attempts have been made
to explain the DIG by photoionization models (e.g. Domg\"orgen \& Mathis
\cite{domma}), while few studies address the possibility of shock excitation
(e.g. Sivan et al. \cite{sivan}). In recent years, DIG was found not only in the
disc of galaxies, but also far above the disc in the halo of the Milky Way and
in some edge-on galaxies at scale heights of more than 1\,kpc (e.g. Reynolds
\cite{rey}; Rand et al. \cite{rand}). Questions therefore arise about where this
extraplanar DIG (eDIG) comes from and how it is ionized. Dynamical models of
galaxies like `galactic fountains' (Shapiro \& Field \cite{shap}) and `chimneys'
(Norman \& Ikeuchi \cite{nor}) describe how gas can be transported from the disc
into the halo. Supernova explosions heating the gas in the disc and pushing it
up into the halo are important for both the dynamics and the ionization of the
gas in the halo. Due to the high velocities in this ejected gas, shocks can
arise and ionize the gas far above the disc. The model of runaway O stars
leaving the disc and moving into the halo (e.g. Gies \cite{gies}) as well as the
theory of photons created by neutrino decay (Sciama \cite{scia}) are further
attempts to explain the ionization of extraplanar DIG.

To obtain more information about the physical properties of the DIG,
spectroscopic data are needed. Up to now spectra of DIG were taken of the
\object{Milky Way} (MW), NGC\,891, NGC\,4631, \object{NGC\,2188} and the
\object{Large Magellanic Cloud} (LMC). A small sample of dwarf galaxies (Martin
\cite{mart}) and a sample of five late-type galaxies (Wang et al. \cite{wang})
have also been examined spectroscopically. In this paper we present
spectroscopic data of the central region of NGC\,55, a highly inclined
($i\sim 77^\circ$), nearby (2\,Mpc) SB(s)m-galaxy of the Sculptor group. Since
late-type galaxies in general are rich in OB stars and bigger \ion{H}{ii}
regions (Kennicutt et al. \cite{keh}), NGC\,55, a late-type near edge-on galaxy,
is ideal for examining the DIG with regard to the ionization process.

Due to its vicinity to the Milky Way, high spatial resolution allows us to
distinguish between different morphological structures of the diffuse gas which
are identified with the help of an H$\alpha$ mosaic of NGC\,55. 

\section{Observations and data reduction}

\subsection{Observations}

\begin{figure*}
\resizebox{\hsize}{!}{\includegraphics{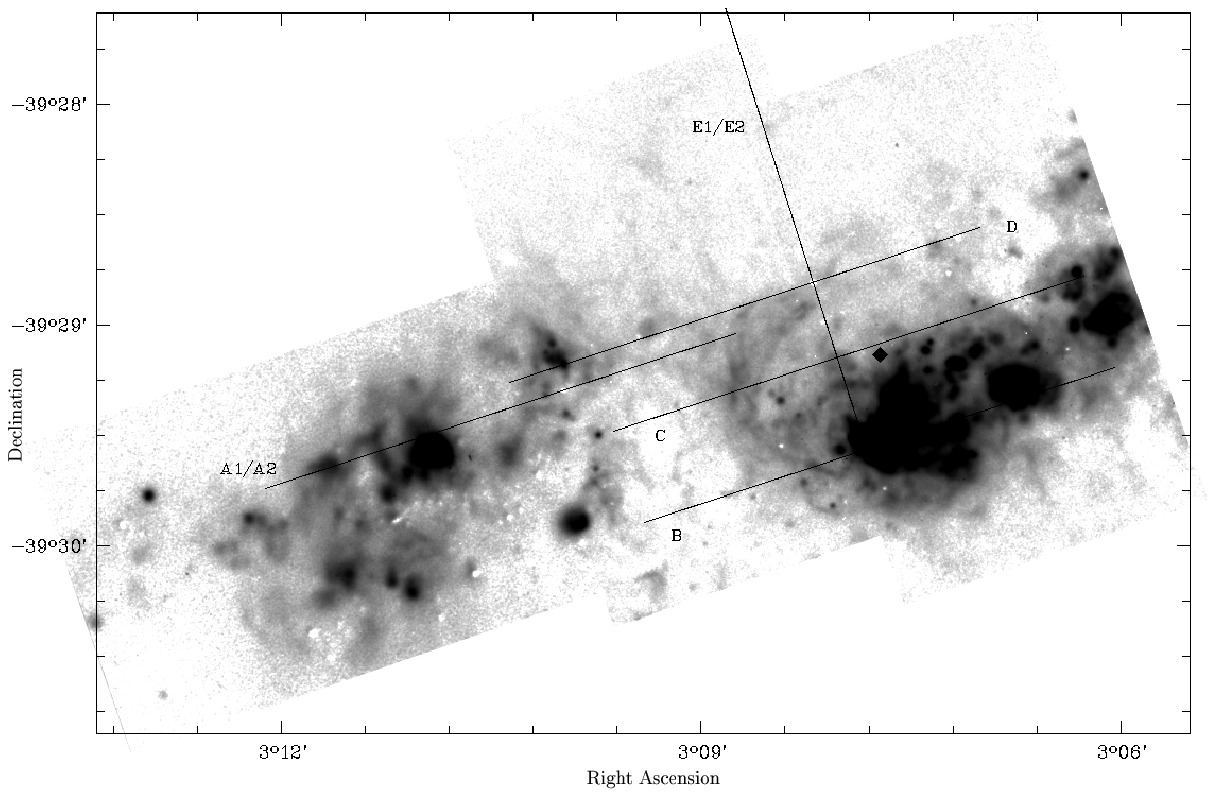}}
\caption{H$\alpha$ image of the central region of NGC\,55. The intensity scale
is logarithmic. Slit positions and the center of rotation (lozenge) are shown.}
\label{pn55}
\end{figure*}
The image of NGC\,55 is a mosaic of three different images obtained at the
2.2\,m telescope at ESO/La Silla. The RCA 5264--16 CCD chip that was used
yielded a pixel size of $0\farcs35$. The field of view of each image was about
$1\farcm9$ x $3\farcm0$. The exposure time was 30\,min for each H$\alpha$ image
(filter \#387, $\lambda_{\rm c}$=6569\,\AA, FWHM=81\,\AA) and 15\,min for each
R-band image (filter \#464, $\lambda_{\rm c}$=6683\,\AA, FWHM=1083\,\AA) used
for continuum subtraction. The filter transmission was the same for the
nitrogen lines ([\ion{N}{ii}] $\lambda\lambda$6548.1, 6583.4\AA) and the
H$\alpha$ line ($\lambda$6562.8\AA).

The spectra of NGC\,55 were obtained at the 3.6\,m telescope at ESO/La Silla.
The B\&C spectrograph with a GEC 2075--11--11 CCD chip yielded a pixel size of
$0\farcs807$ and 1.28\,\AA, respectively. The slit length was about $2\farcm7$
in a wavelength range of 6258--6987\,\AA. The exposure time for each spectrum
was 30\,min.

\subsection{Data reduction}

All data for NGC\,55 were reduced by standard procedures in MIDAS. Since the
emission lines in the spectra covered the entire slit, it was not possible to
use sky spectra from the same frames. Therefore sky spectra obtained from other
galaxy spectra observed during the same night at the same telescope were
compared with each other. It turned out that the line ratios among the sky lines
remained constant during the night. Thus we could use these sky spectra for sky
subtraction in the NGC\,55 data. The correction of the spectroscopic data with
regard to the response function of the CCD was negligible. To calculate line
ratios and velocities, we took means over several rows (if necessary) before
measuring the line position and its count rate by fitting a Gaussian curve to
the line profile. Taking the averages we carefully prevented the smoothing of
variations of possible astrophysical origin. Coordinates refer to equinox
B1950.0.

\section{Results and discussion}

\begin{figure*}
\resizebox{\hsize}{!}{\includegraphics{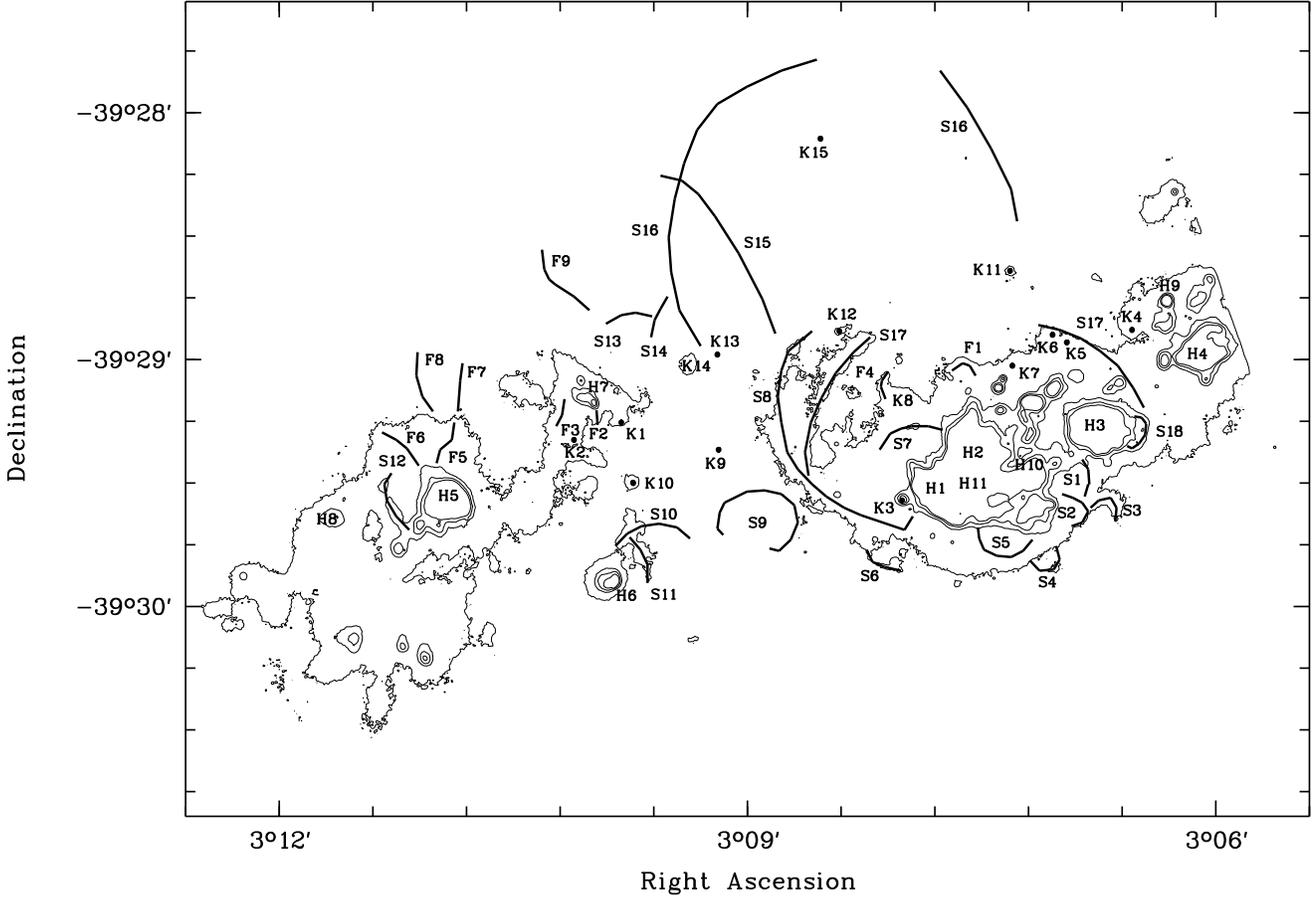}}
\caption{Contour plot of the central region of NGC\,55. The structures are
named by capital letters as follows: \ion{H}{ii} region (H), shell (S), knot (=
brighter, compact region of DIG, K), filament (F)}
\label{pstruk55}
\end{figure*}

In Fig.~\ref{pn55} the central region of NGC\,55 in H$\alpha$ and the slit
positions of the spectra are shown. The slits are parallel and perpendicular,
respectively, to the disc. The center of rotation (lozenge in Fig.~\ref{pn55})
was determined by \ion{H}{i} measurements by Puche et al. (\cite{puch}). All
structures identified and named by capital letters (H = \ion{H}{ii} region, S =
shell, F = filament, K = knot) are shown in Fig.~\ref{pstruk55}. We examined
most of these structures spectroscopically.

\subsection{Distribution of ionized gas}

The integrated flux in H$\alpha$ and NII is determined to be
2.8$\times$10$^{-11}$\,erg\,sec$^{-1}$\,cm$^{-2}$. The 3$\sigma$ detection limit
in the H$\alpha$+[NII] image is 1.5$\times$10$^{-17}$\,erg\,sec$^{-1}$\,cm$^
{-2}$/$\sq\arcsec$. Since the [NII] contribution is small (see Table \ref{ira}),
fluxes can be directly converted into emission measures (EM). Faint structures
in Fig.~\ref{pn55} such as S16 are at this detection limit of
EM$\approx$20\,pc\,cm$^{-6}$, while the grey scale saturates at a level of
EM=1100\,pc\,cm$^{-6}$.

Several large  \ion{H}{ii} regions are clearly visible in Fig.~\ref{pn55}. Many
shell-like structures seem to arise out of these \ion{H}{ii} regions. The most
striking features are the loop S8 with an extension of 540\,pc (a distance of
2\,Mpc is assumed throughout this paper) associated with the \ion{H}{ii} complex
H1/H2 and the superbubble S16. This bubble was discovered first by Graham \&
Lawrie (\cite{grala}) in [\ion{O}{iii}] $\lambda$5007\AA\ narrow band images.
Corrected for the different distances used, the diameter of this superbubble was
estimated by Graham \& Lawrie to be 920\,pc, which is similar to 960\,pc
determined by Ferguson et al. (\cite{fer}). However, with our better spatial
resolution we estimate its extension to be about 1270\,pc. Our data have errors
of $\pm$30--40\,pc for larger features ($>$130\,pc). Another interesting shell
is S17 with a radius of 410--520\,pc. Most of the brightest \ion{H}{ii} regions
seem to lie inside this shell. This structure, however, is not closed; there is
no boundary visible to the north of H1/H2. Diffuse gas seems to gush out of the
shell as from a bursting balloon. The structures identified in
Fig.~\ref{pstruk55} verify morphologically the model of diffuse gas ionized by
\ion{H}{ii} regions and pushed out into the halo by supernova explosions.

Nevertheless, not every shell-like structure has a corresponding \ion{H}{ii}
region, as the almost complete ring S9 shows. With a diameter of 170--230\,pc
this ring is too large to be a supernova remnant. One possible explanation for
this feature is a source with a much shorter lifetime than the shell itself.
Whether an ionizing source is hidden by dust can be checked in the future by NIR
observations. Shells without visible source are also observed in other galaxies.
However, it is still unknown, how these structures are formed and what their
sources are (Hunter et al. \cite{hhg}).

\subsection{Spectroscopic data}

After examining the structures of Fig.~\ref{pn55} and Fig.~\ref{pstruk55},
respectively, we were able to identify most of them in the line profiles of our
spectra. We calculated their line ratios, heliocentric velocities and their
intrinsic line widths. We only used data with a signal-to-noise ratio of
S/N$>$3. Due to the interaction between NGC\,55 and two neighbouring galaxies,
the disc of NGC\,55 is warped (Hummel et al. \cite{humm}). Therefore its
inclination is very uncertain, and it was impossible to calculate the exact
distance of each structure to the midplane. 

\subsubsection{Relative velocities}

\begin{figure}
\resizebox{\hsize}{!}{\includegraphics{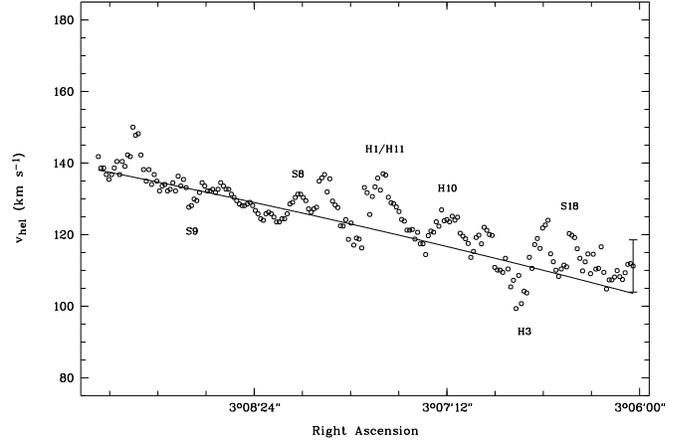}}
\caption{Heliocentric velocities of the H$\alpha$ line along slit B. The
estimated error is about $\pm$7\,$\rm km\,s^{-1}$ (error bar on the right). For
comparison the rotation curve (i.e. rotation velocity plus system velocity)
fitted to \ion{H}{i} measurements (Puche et al. \cite{puch}) is also plotted.}
\label{pvhel55b}
\end{figure}

The emission lines of each spectrum yielded identical recession velocities along
the slit within the errors of $\sim$0.16\,\AA\ ($\pm$7\,$\rm km\,s^{-1}$ for
H$\alpha$). Since H$\alpha$ was the brightest line in each spectrum and could be
traced along the entire slit, we used this line to get heliocentric recession
velocities. These velocities were compared with the rotation curve (i.e.
rotation velocity plus system velocity) fitted to \ion{H}{i} measurements (Puche
et al. \cite{puch}). In Fig.~\ref{pvhel55b} this comparison is shown for slit B.
In all slits the DIG mostly corotates with the galaxy. Only the velocity of the
diffuse gas in the vicinity of the shells S17 and S8 differs from galactic
rotation. Thus, dynamics seems to play an important role in this area. To get
velocities relative to the vicinity of each feature, we decided individually
whether to use the rotation curve or the majority of the \ion{H}{ii} motion as
ambient velocity. Most of the \ion{H}{ii} regions -- fainter regions as well as
brighter regions -- have velocities of about 12\,$\rm km\,s^{-1}$ relative to
the \ion{H}{i}. These velocities are observed both as corrotation and
antirotation. Unfortunately we do not have enough data to determine whether this
result has a physical cause or whether it is just a random effect. However, most
of the shells have relative velocities of 5--10\,$\rm km\,s^{-1}$, whereas the
DIG above the loop S8 inside the superbubble S16 is strongly blue shifted with
relative velocities of more than 40\,$\rm km\,s^{-1}$.

\subsubsection{Line ratios}

In the following we list for doublets the brighter component only, i.e.
[\ion{S}{ii}] $\lambda$6716.4\AA, [\ion{N}{ii}] $\lambda$6583.4\AA\ and
[\ion{O}{i}] $\lambda$6300.3\AA.

\begin{table}
\caption[]{\label{ira55tab}Maximal extensions of shell structures (in pc) and
measured line ratios for all features.}
\begin{tabular}{lllllll} \hline\noalign{\smallskip}
Name & Slit & Radius & $\frac{\rm [\ion{O}{i}]}{\rm H\alpha}$ & $\frac{\rm
[\ion{N}{ii}]}{\rm H\alpha}$ & $\frac{\rm \ion{He}{i}}{\rm H\alpha}$ & $\frac
{\rm [\ion{S}{ii}]}{\rm H\alpha}$ \\
\noalign{\smallskip}\hline\noalign{\smallskip}
H1 & B & & 0.0052 & 0.036 & 0.011 & 0.075 \\
H1 & E2 & & 0.0055 & 0.034 & 0.01 & 0.049 \\
H3 & B & & 0.0055 & 0.027 & 0.01 & 0.068 \\
H5 & A2 & & 0.004 & 0.039 & 0.011 & 0.063 \\
H7 & D & & 0.016 & 0.057 & & 0.103 \\
H8 & A2 & & & 0.057 & & 0.09 \\
H9 & C & & & & & 0.095 \\
H10 & B & & 0.0083 & 0.058 & 0.008 & 0.115 \\
H11 & B & & 0.004 & 0.026 & 0.01 & 0.068 \\ \noalign{\smallskip}
K1 & A2 & & & 0.08 & & 0.21 \\
K2 & A2 & & & 0.06 & & 0.1 \\
K3 & B & & 0.015 & 0.08 & & 0.16 \\
K4 & C & & 0.03 & 0.12 & & 0.21 \\
K5 & C & & 0.018 & 0.1 & & 0.19 \\
K6 & C & & 0.015 & 0.11 & & 0.2 \\
K7 & C & & 0.039 & 0.09 & & 0.21 \\
K8 & C & & 0.042 & 0.09 & & 0.2 \\
K8 & E2 & & 0.038 & 0.07 & & 0.16 \\
K11 & D & & & 0.07 & & 0.17 \\
K12 & D & & & 0.06 & & 0.09 \\
K13 & D & & & 0.08 & & 0.19 \\
K14 & D & & & 0.07 & & 0.22 \\ \noalign{\smallskip}
S1 & & 80 & & & & \\
S2 & & 100 & & & & \\
S3 & & 70 & & & & \\
S4 & & 100 & & & & \\
S5 & & 100 & & & & \\
S6 & & 130 & & & & \\
S7 & E2 & 200 & 0.044 & 0.06 & & 0.11 \\
S8 & B & 540 & & 0.08 & & 0.19 \\
S8 & C & 540 & & 0.09 & & 0.16 \\
S8 & D & 540 & & 0.06 & & 0.11 \\
S9 & & $^\ast$230 & & & & \\
S10 & & 150 & & & & \\
S11 & & 120 & & & & \\
S12 & A2 & 150 & 0.013 & 0.06 & 0.01 & 0.12 \\
S13 & & 160 & & & & \\
S14 & & 170 & & & & \\
S15 & A2 & $\sim$660 & & 0.06 & & 0.13 \\
S15 & D & $\sim$660 & & 0.08 & & 0.2 \\
S16 & A2 & 1270 & & 0.06 & & 0.13 \\
S16 & D & 1270 & & 0.09 & & 0.21 \\
S17 (E) & C & 520 & & 0.06 & & 0.13 \\
S17 (W) & C & 520 & 0.026 & 0.09 & & 0.17 \\
S17 & E2 & 520 & & & & 0.13 \\ \noalign{\smallskip}
F1 & C & & 0.043 & 0.08 & & 0.19 \\ \noalign{\smallskip}\hline
\end{tabular} \\
Note: $^\ast$ = diameter (in pc)
\end{table}
\begin{figure}
\resizebox{\hsize}{!}{\includegraphics{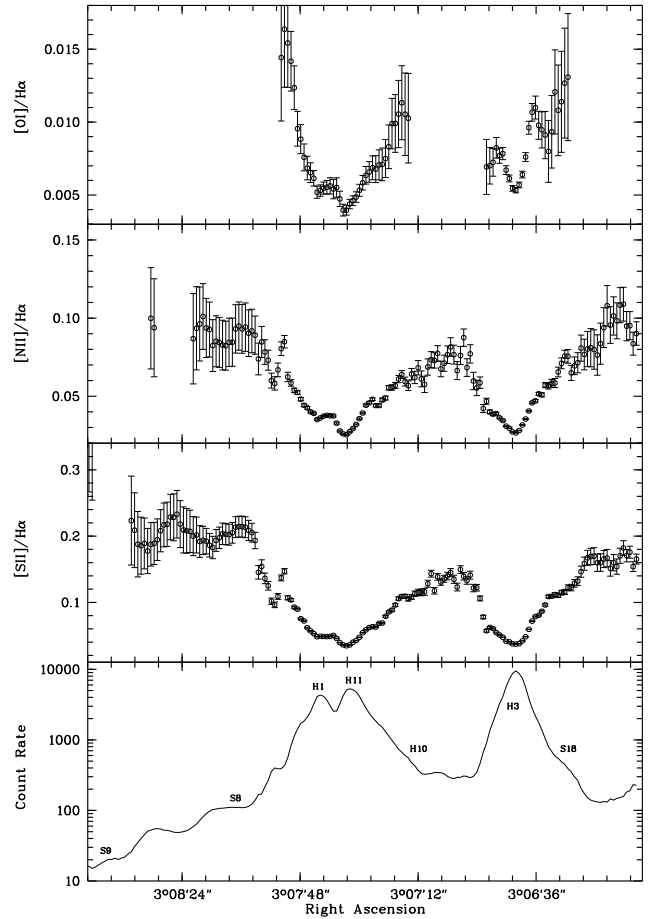}}
\caption{The line ratios [\ion{O}{i}]/H$\alpha$, [\ion{N}{ii}]/H$\alpha$ and
[\ion{S}{ii}]/H$\alpha$ and the count rate of the H$\alpha$ line along slit B.}
\label{pira55b}
\end{figure}
In Fig.~\ref{pira55b} we show the different line ratios of slit B and the count
rate of the H$\alpha$ line. The run of the H$\alpha$ intensity clearly
demonstrates the contrast between \ion{H}{ii} regions and the surrounding
diffuse gas. If line ratios are compared to the H$\alpha$ intensities, the same
general behaviour can be observed along all slits: the larger the count rate or
emission measure is, the lower are the line ratios. This was already shown for
NGC\,55 by Ferguson et al. (\cite{fer}) in their Fig.~3, and it is also true
when plotted against other lines, e.g. [\ion{S}{ii}] and [\ion{N}{ii}] (although
we observe no gap between \ion{H}{ii} regions and DIG like in Ferguson et al.).
The line ratios reach a kind of saturation with increasing distance from the
\ion{H}{ii} regions. This can be observed for both line ratios of different
emission lines (as shown in Fig.~\ref{pira55b}) and line ratios of different
spectra. 

Under the assumption of radiation bounded \ion{H}{ii} regions, no ionization of
diffuse gas by OB stars would be expected, and thus another ionization process
has to be responsible for the H$\alpha$ emission of the diffuse gas. If the
\ion{H}{ii} regions are density bounded, photons of the OB stars can still reach
the diffuse gas surrounding the \ion{H}{ii} regions. In this case, the photon
field is very diluted and yields higher line ratios that do not increase with
increasing distance of the \ion{H}{ii} regions. This behaviour is observed in
all line ratios of our spectra. Smaller variations in the line ratios can be
explained by slight changes in the density and therefore in the opacity of the
DIG, since the ionization parameter of photoionization models depends on the
density of the gas (Domg\"orgen \& Mathis \cite{domma}).

\begin{figure*}
\resizebox{12cm}{!}{\includegraphics{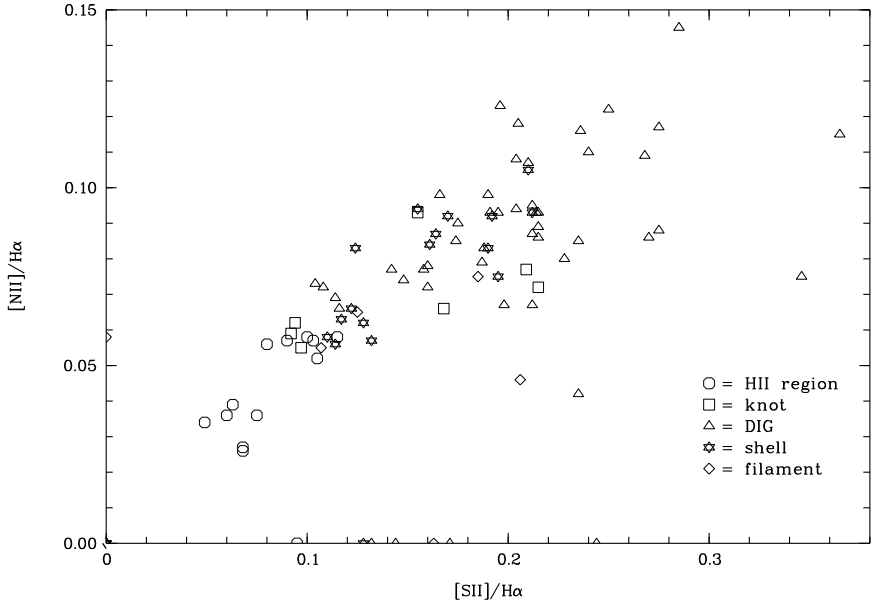}}
\hfill
\parbox[b]{55mm}{
\caption{Comparison of the line ratios [\ion{N}{ii}]/H$\alpha$ and
[\ion{S}{ii}]/H$\alpha$ of the identified structures of all slits. The errors
increase with increasing line ratios ($\pm$0.001--0.016 for
[\ion{S}{ii}]/H$\alpha$ and $\pm$0.001--0.017 for [\ion{N}{ii}]/H$\alpha$).}
\label{pns55}}
\end{figure*}
In Fig.~\ref{pns55} we compare the line ratios [\ion{N}{ii}]/H$\alpha$ and
[\ion{S}{ii}]/H$\alpha$ with each other distinguishing between different
features identified in the spectra. It is clearly visible that \ion{H}{ii}
regions have lower line ratios than the diffuse ionized gas. The line ratios of
shell structures are in between and have a large spread reaching almost the
values of the \ion{H}{ii} regions and the lower values of the DIG. No difference
in the line ratios could be established for DIG in the disc and DIG in the halo.
The ratio [\ion{N}{ii}]/[\ion{S}{ii}] and thus the slope in Fig.~5 is almost
constant ($\sim$0.4). Therefore this ratio seems to be independent of the
excitation process of nitrogen and sulfur (Goad \& Roberts \cite{goad}). In
Table \ref{ira55tab} the line ratios of the different structures and the radii
of the shells are listed.

While it is not possible to derive the N/S ratio from the
[\ion{N}{ii}]/[\ion{S}{ii}] ratio, changes in the ratio
[\ion{N}{ii}]/[\ion{S}{ii}] can still partly reflect variations in the
abundances of nitrogen and sulfur (Hawley \& Grandi \cite{hagra}). It is
frequently argued that nitrogen is a secondary product of nucleosynthesis
produced by stars of about 1\,M$_\odot$ developed from metal enriched clouds. On
the contrary, sulfur is produced by massive stars of about 25\,M$_\odot$
developed from metal poor clouds (Rubin et al. \cite{rubin}). Since late-type
galaxies contain more massive stars than early-type galaxies (Kennicutt et al.
\cite{keh}), the ratio N/S should be lower in late-type galaxies, and so should
be the ratio [\ion{N}{ii}]/[\ion{S}{ii}]. We summarized the
[\ion{N}{ii}]/H$\alpha$ and [\ion{S}{ii}]/H$\alpha$ line ratios in Table
\ref{ira} and compared them with line ratios of other galaxies. We measured
somewhat lower line ratios for NGC\,55 than Ferguson et al. (\cite{fer}) (they
found [\ion{S}{ii}] line ratios of 0.05--0.3 for \ion{H}{ii} regions and
0.2--0.7 for DIG). This can easily be explained by the fact that Ferguson et al.
calculated [\ion{S}{ii}]$\lambda\lambda$6716.4,6730.8\AA/H$\alpha$+[\ion{N}
{ii}]$\lambda\lambda$6548.1,6583.4\AA\ and that our line ratios are measured in
the central region (with a denser photon field), whereas Ferguson et al. took
mean values of the entire galaxy. However, as one can see in Table \ref{ira},
most late-type galaxies (Sd, Sm) have lower [\ion{N}{ii}]/[\ion{S}{ii}] ratios
than the early-type galaxies (Sa, Sb) -- as predicted by the considerations
above and confirmed by the line ratios given by Wang et al. (\cite{wang}) (their
Table 6). Due to the fact that NGC\,55 has a relatively low star formation rate
(0.3\,M$_\odot$\,yr$^{-1}$, Dettmar \& Heithausen \cite{dehei}) and that
late-type galaxies and irregular galaxies had a rather constant star formation
rate over a longer period (Kennicutt \cite{kenn}), the nitrogen abundance
decreases even more in NGC\,55 and thus the ratio [\ion{N}{ii}]/[\ion{S}{ii}].
How sensitive this ratio is to variations in the nitrogen and sulfur abundances,
is shown by model calculations by Domg\"orgen \& Mathis (\cite{domma}).

\begin{table*}
\caption[]{\label{ira}Comparison of the line ratios [\ion{S}{ii}]/H$\alpha$ and
[\ion{N}{ii}]/H$\alpha$ for NGC\,55 with line ratios of other galaxies.}
\begin{tabular}{lllllllll} \hline\noalign{\smallskip}
Galaxy & Type & \multicolumn{2}{c}{[\ion{S}{ii}]/H$\alpha$} & \ &
\multicolumn{2}{c}{[\ion{N}{ii}]/H$\alpha$} & \ & [\ion{N}{ii}]/[\ion{S}{ii}] \\
\cline{3-4}\cline{6-7}
& & min. & max. & & min. & max. & & \\ \noalign{\smallskip}\hline
\noalign{\smallskip}
\object{NGC\,55}, \ion{H}{ii} & SB(s)m & 0.05$\pm$0.01 & 0.11$\pm$0.01 & &
0.025$\pm$0.001 & 0.056$\pm$0.005 & & \\
NGC\,55, DIG & & 0.12$\pm$0.02 & 0.28$\pm$0.01 & & 0.065$\pm$0.010 &
0.12$\pm$0.01 & & 0.4 \\
NGC\,55, Shells & & 0.11 & 0.21 & & 0.055 & 0.095 & & \\ \noalign{\smallskip}
\hline\noalign{\smallskip}
\object{IC\,4662}, \ion{H}{ii}\,$^1$ & IBm & 0.080$\pm$0.005 & 0.19$\pm$0.01 & &
0.035$\pm$0.004 & 0.075$\pm$0.005 & & \\
IC\,4662, DIG\,$^1$ & & 0.105$\pm$0.010 & 0.22$\pm$0.01 & & 0.055$\pm$0.005 &
0.105$\pm$0.005 & & 0.5 \\
IC\,4662, Shells\,$^1$ & & 0.12 & 0.17 & & 0.058 & 0.062 & & \\
\noalign{\smallskip}\hline\noalign{\smallskip}
\object{IC\,5052}, \ion{H}{ii}\,$^1$ & SBd & 0.030$\pm$0.002 & 0.080$\pm$0.005 &
& 0.025$\pm$0.002 & 0.08$\pm$0.01 & & 1.2 \\
IC\,5052, DIG\,$^1$ & & 0.12$\pm$0.02 & 0.17$\pm$0.02 & & 0.10$\pm$0.01 &
0.135$\pm$0.010 & & \\ \noalign{\smallskip}\hline\noalign{\smallskip}
\object{LMC}, \ion{H}{ii}\,$^2$ & & 0.04 & 0.49 & & 0.02 & 0.13 & & 0.3--0.5 \\
LMC, Shells\,$^2$ & & 0.06 & 0.84 & & 0.03 & 0.24 & & \\ \noalign{\smallskip}
\hline\noalign{\smallskip}
\object{NGC\,4631}, \ion{H}{ii}\,$^3$ & SB(s)d & 0.12 & 0.13 & & 0.09 & 0.17 & &
0.5--1.3 \\
NGC\,4631, DIG\,$^3$ & & 0.36 & 0.50 & & 0.18 & 0.49 & & \\ \noalign{\smallskip}
\hline\noalign{\smallskip}
\object{NGC\,891}, \ion{H}{ii}\,$^4$ & SA(s)b & 0.1 & & & 0.3 & 0.4 & & 1.2--2.4
\\
NGC\,891, DIG\,$^4$ & & 0.25 & 0.60 & & 0.4 & 1.1 & & \\ \noalign{\smallskip}
\hline\noalign{\smallskip}
\object{MW}, \ion{H}{ii}\,$^4$ & & $\sim$0.1 & & & $\sim$0.3 & & & $>$1.0 \\
MW, DIG\,$^4$ & & 0.3 & 0.5 & & 0.3 & 0.5 & & \\ \noalign{\smallskip}\hline
\end{tabular} \\
References: $^1$ = Otte \cite{otte}, $^2$ = Hunter \cite{hunt}, $^3$ = Golla et
al. \cite{golla}, $^4$ = Dettmar \cite{det}
\end{table*}

The low metallicity in late-type galaxies has two different effects on the line
ratios. Since forbidden lines of metals are important cooling lines (Aller
\cite{all}), the gas has a higher temperature in galaxies of low metallicity,
thereby increasing the strength of collisionally excited lines over the Balmer
recombination lines. On the other hand low metal abundance leads to a hardening
of the ionizing radiation due to less absorption in the UV and results in a
higher ionization for low ionization species. Since we observe lower line ratios
for metal poor galaxies, photoionization seems to be again the most important
ionization process. The [\ion{N}{ii}]/[\ion{S}{ii}] however might also be
affected by geometrical effects due to the very different ionization levels for
these two species as discussed by Petuchowski \& Bennett (\cite{pet1},
\cite{pet2}). 

\begin{figure*}
\resizebox{12cm}{!}{\includegraphics{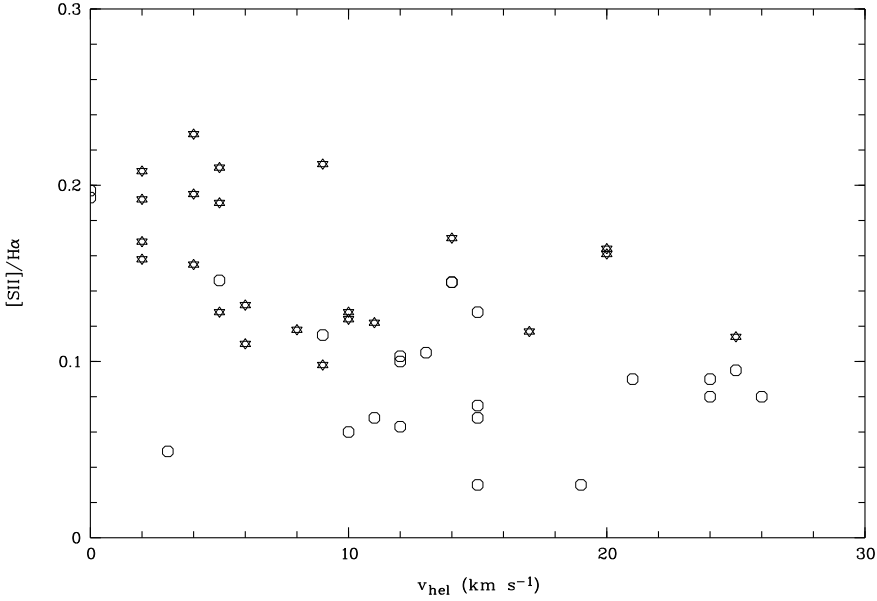}}
\hfill
\parbox[b]{55mm}{
\caption{Comparison between the [\ion{S}{ii}]/H$\alpha$ line ratio and the
relative velocity of the shell structures (stars) and the \ion{H}{ii} regions
(circles).}
\label{piv55}}
\end{figure*}

Due to the corotation of the eDIG with the disc and the observed shell
structures reaching into the halo, the eDIG seems to be pushed out of the disc
into the halo by supernova explosions. Therefore the question arises whether the
diffuse gas can be ionized by shocks. Shock ionization produces much higher line
ratios than photoionization (e.g. Peimbert et al. \cite{peim}). The higher the
expansion velocity of a shell is, the higher should be the line ratios due to
shockionization. In Fig.~\ref{piv55} we compare the measured [\ion{S}{ii}] line
ratios of shells and \ion{H}{ii} regions with their relative velocities. The
trend for shells is more or less the reverse of what would be expected due to
shockionization. The \ion{H}{ii} regions show almost constant
[\ion{S}{ii}]/H$\alpha$ ratios independent of their relative velocities. Thus,
it seems to be unlikely that shockionization plays an important role in NGC\,55.

Martin (\cite{mart}) showed for her sample of 14 dwarf galaxies that
photoionization of leaking \ion{H}{ii} regions fits the lower line ratios,
whereas additional emission from shock-excited gas is needed to explain the
relatively high line ratios measured in lower surface brightness regions. Thus,
it is still possible that we are missing very faint high velocity shock excited
gas.

\begin{figure*}
\resizebox{12cm}{!}{\includegraphics{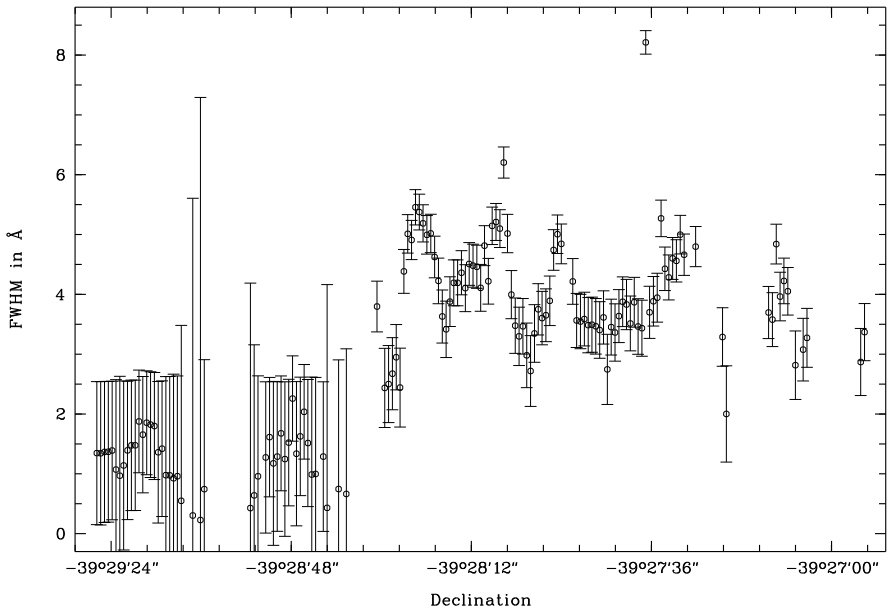}}
\hfill
\parbox[b]{55mm}{
\caption{The intrinsic line width (FWHM) of the H$\alpha$ line of slit E2. It
increases rapidly from 1\,\AA\ up to 4.25\,\AA\ at the same point where the
relative velocity increases to more than 40\,$\rm km\,s^{-1}$.}
\label{pfwhm55e2}}
\end{figure*}

\subsubsection{The superbubble}

An interesting feature is the superbubble S16. Its line ratios in the lower part
near the disc resemble the higher line ratios of the DIG (compare Tables
\ref{ira55tab} and \ref{ira}). Unfortunately, the emission lines in the slits
E1/E2 were too weak to calculate line ratios for the upper part of S16 at
distances of more than 1\,kpc above the disc. Nevertheless, it was possible to
measure the H$\alpha$ line at this height. As mentioned above, the DIG inside
this bubble is blue shifted with relative velocities of more than
40\,$\rm km\,s^{-1}$. At the same point where the relative velocity increases,
the intrinsic line width (FWHM) increases rapidly from 1\,\AA\ up to about
4.25\,\AA\ (Fig.~\ref{pfwhm55e2}). The intrinsic line width generally is about
1\,\AA\ along all other slits. A FWHM of 4.25\,\AA\ corresponds to a velocity
distribution of about 194\,$\rm km\,s^{-1}$ and therefore to an expansion
velocity of at least 100\,$\rm km\,s^{-1}$ for the shell S16. The highest
expansion velocity measured in the LMC is 68\,$\rm km\,s^{-1}$ (Hunter
\cite{hunt}).

Assuming that the density inside the superbubble S16 is low enough, photons of
the OB stars in the \ion{H}{ii} regions can ionize the diffuse gas high above
the disc at the top of S16. This is valid also for the other shells observed.

Although NGC\,55 and the LMC are very similar in most of their properties, they
show slightly different properties in their shell populations. While the extent
of LMC shells lies in the range of 35--1400\,pc and their line ratios are
0.03--0.24 ([\ion{N}{ii}]/H$\alpha$) and 0.06--0.84 ([\ion{S}{ii}]/H$\alpha$),
the ranges in NGC\,55 are smaller for both the expansions of the shells
(80--1270\,pc) and the line ratios ([\ion{N}{ii}]/H$\alpha$=0.055--0.095,
[\ion{S}{ii}]/H$\alpha$=0.10--0.21). Hunter (\cite{hunt}) showed for the LMC,
that the diffuse gas is ionized by photons of OB stars instead of shocks.
However, one should note that this study refers to brighter DIG only.

\section{Summary}

In this paper we presented relative velocities and emission line ratios for
various structures in the diffuse ionized gas of the late-type galaxy NGC\,55
and compared these properties with each other and with properties of other
galaxies. The essential results are summarized in the following.
\begin{itemize}
\item Most of the \ion{H}{ii} regions observed in NGC\,55 have relative
velocities of about 12\,$\rm km\,s^{-1}$ with regard to the ambient rotation
velocity of NGC\,55. The reason for this behaviour cannot be determined without
more data. The in-plane shell structures have typical relative velocities of
about 5--10\,$\rm km\,s^{-1}$.
\item The observed kinematic and the shell structures indicate, that diffuse gas
is expelled into the halo by supernova explosions or stellar winds. 
\item Not every shell has a visible ionization source. Probably these shells
have sources of shorter lifetimes. But the question remains, how these shells
can stay in ionization equilibrium without visible ionizing sources.
\item Shockionization cannot play an important role in the ionization processes
of the diffuse gas. The low metallicity, the coherence between the [\ion{S}{ii}]
line ratios and the relative velocities, the low line ratios themselves, the
similarity between NGC\,55 and the LMC, all indicate that photoionization is the
most important ionization process in NGC\,55.
\item The diffuse ionized gas in the disc and the eDIG do not differ from each
other with regard to their line ratios. We have not found any evidence that it
would be necessary to distinguish between a ``quiescent'' DIG and a
``disturbed'' DIG as Wang et al. (\cite{wang}) did.
\item There seems to be a relationship between the Hubble-type of a galaxy and
its line ratios. Late-type galaxies and irregular galaxies have lower line
ratios than early-type galaxies. The [\ion{N}{ii}]/[\ion{S}{ii}] ratio is lower
for late-type galaxies.  Both are in accord with expectations for low
metallicity environments.

\end{itemize}

\begin{acknowledgements}
We thank R. J. Reynolds, J. S. Gallagher, D. Bomans and M. Rosa for their
comments on early drafts of this paper. We also thank the referee J.-P. Sivan
for his suggestions and comments which improved the paper. B. O. is thankful to
her colleagues in the Department of Astronomy for their friendly support while
writing this paper.
\end{acknowledgements}

\end{document}